\documentclass[sn-mathphys-num,iicol]{sn-jnl}

\usepackage{lineno}

\usepackage{graphicx}%
\usepackage{multirow}%
\usepackage{amsmath,amssymb,amsfonts}%
\usepackage{amsthm}%
\usepackage{mathrsfs}%
\usepackage[title]{appendix}%
\usepackage{xcolor}%
\usepackage{textcomp}%
\usepackage{manyfoot}%
\usepackage{booktabs}%
\usepackage{algorithm}%
\usepackage{algorithmicx}%
\usepackage{algpseudocode}%
\usepackage{listings}%

\usepackage{bm}

\theoremstyle{thmstyleone}%
\theoremstyle{thmstyletwo}%

\theoremstyle{thmstylethree}%

\begin{document}

\title[Article Title]{Modality-Projection Universal Model for Comprehensive Full-Body Medical Imaging Segmentation}


\author[1]{\fnm{Yixin} \sur{Chen}}
\equalcont{These authors contributed equally to this work.}

\author[2]{\fnm{Lin} \sur{Gao}}
\equalcont{These authors contributed equally to this work.}

\author[3]{\fnm{Yajuan} \sur{Gao}}

\author[4]{\fnm{Rui} \sur{Wang}}

\author[3]{\fnm{Jingge} \sur{Lian}}

\author[5]{\fnm{Xiangxi} \sur{Meng}}

\author[2]{\fnm{Yanhua} \sur{Duan}}
\author[2]{\fnm{Leiying} \sur{Chai}}

\author[1,3,6]{\fnm{Hongbin} \sur{Han}}

\author*[2]{\fnm{Zhaoping} \sur{Cheng}}\email{czpabc@163.com}

\author*[1]{\fnm{Zhaoheng} \sur{Xie}}\email{xiezhaoheng@pku.edu.cn}



\affil[1]{\orgdiv{the Institute of Medical Technology and National Biomedical Imaging Center}, \orgname{Peking University}, \orgaddress{\city{Beijing}, \postcode{100191},  \country{China}}}

\affil[2]{\orgdiv{Department of Nuclear Medicine}, \orgname{The First Affiliated Hospital of Shandong First Medical University \& Shandong Provincial Qianfoshan Hospital}, \orgaddress{\city{Jinan}, \postcode{250014},  \country{China}}}

\affil[3]{\orgdiv{Department of Radiology}, \orgname{Peking University Third Hospital}, \orgaddress{\city{Beijing}, \postcode{100191},  \country{China}}}

\affil[4]{\orgdiv{Department of Radiology}, \orgname{Guangdong Provincial People's Hospital}, \orgaddress{\city{Guangdong}, \country{China}}}

\affil[5]{\orgdiv{Key Laboratory of Carcinogenesis and Translational Research (Ministry of Education/Beijing)}, \orgdiv{Key Laboratory for Research and Evaluation of Radiopharmaceuticals (National Medical Products Administration)}, \orgdiv{Department of Nuclear Medicine}, \orgname{Peking University Cancer Hospital \& Institute}, \orgaddress{\city{Beijing}, \postcode{100142},  \country{China}}}

\affil[6]{\orgdiv{Beijing Key Laboratory of Magnetic Resonance Imaging Devices and Technology}, \orgname{Peking University Third Hospital}, \orgaddress{\city{Beijing}, \postcode{100191},  \country{China}}}


\abstract{The integration of deep learning in medical imaging has shown great promise for enhancing diagnostic, therapeutic, and research outcomes. However, applying universal models across multiple modalities remains challenging due to the inherent variability in data characteristics. This study aims to introduce and evaluate a Modality Projection Universal Model (MPUM). MPUM employs a novel modality-projection strategy, which allows the model to dynamically adjust its parameters to optimize performance across different imaging modalities. The MPUM demonstrated superior accuracy in identifying anatomical structures, enabling precise quantification for improved clinical decision-making. It also identifies metabolic associations within the brain-body axis, advancing research on brain-body physiological correlations. Furthermore, MPUM's unique controller-based convolution layer enables visualization of saliency maps across all network layers, significantly enhancing the model’s interpretability.}




\maketitle

\section{Introduction}\label{sec1}

Universal models, characterized by their ability to generalize across diverse tasks without fine-tuning, have emerged as a powerful framework in many fields. By leveraging shared representations\cite{pcnet,medsam}, these models offer unparalleled adaptability across a wide range of applications. In the realm of medical imaging, universal models have gained attention for their potential to generalize across various anatomical regions, imaging modalities, and clinical tasks. However, significant challenges arise due to the diverse modalities and intricate anatomical structures involved. In response to these challenges, several large-scale datasets, such as TotalSegmentator\cite{totalsegmentator,totalsegmentatormri} and Dense Anatomical Prediction (DAP)\cite{dap}, various universal challenges like BodyMaps24\cite{bodymaps24} and the Universal Lesion Segmentation\cite{uls23}, as well as related universal models like MedSAM\cite{medsam}, CDUM\cite{cdum}, PCNet\cite{pcnet}, TotalSegmentator\cite{totalsegmentator}, STUNet\cite{stunet}, and LUCIDA\cite{lucida}, propelled progress in this domain, demonstrating the immense potential.

In the medical imaging field, universal models possess powerful multi-task processing capabilities, enabling the quick and accurate identification of human tissues. These models assist clinicians by distinguishing different organs, thereby aiding in the diagnostic process. After completing radiological scans and nuclear medicine scan, a universal model can automatically identify all tissue structures, generating preliminary reports and reducing doctors' workload. Additionally, these models can monitor patients undergoing treatment by detecting structural or functional changes, detecting structural or functional changes. They also reduce the time and manpower required for annotating specific tissue regions of interest (ROIs), which is particularly beneficial in clinical research.

Similar to universal models, foundation models represent an alternative approach to developing generalizable AI systems in medical imaging. While foundation models rely on pre-training through self-supervised learning on large amounts of unannotated or weakly annotated data\cite{foundation_biomarker,chen2020simple,caron2020unsupervised,dwibedi2021little}, universal models focus on training with annotated multi-task datasets. Both approaches aim to generate high-dimensional representations that can generalize across various tasks. However, foundation models face inherent limitations, particularly in multi-category segmentation tasks, where the binary nature of contrastive learning is less effective. Additionally, foundation models typically require task-specific fine-tuning, which compromises their practicality as truly versatile tools in medical applications. In contrast, universal models face their own set of challenges. Chief among them is the need for large-scale annotated datasets. Universal models depend heavily on high-quality, labeled data from diverse tasks\cite{upadhyay2024advances,yang2024multi}. This requirement presents a significant bottleneck, as obtaining sufficient annotated data for medical imaging tasks can be labor-intensive. Despite these limitations, researching universal models is crucial because they offer the potential for truly versatile tools in medical applications without the need for task-specific fine-tuning\cite{pcnet}. Universal models can handle multiple tasks simultaneously, leading to more efficient clinical workflows.

\begin{figure*}[t]
    \centering
    \includegraphics[width=\textwidth]{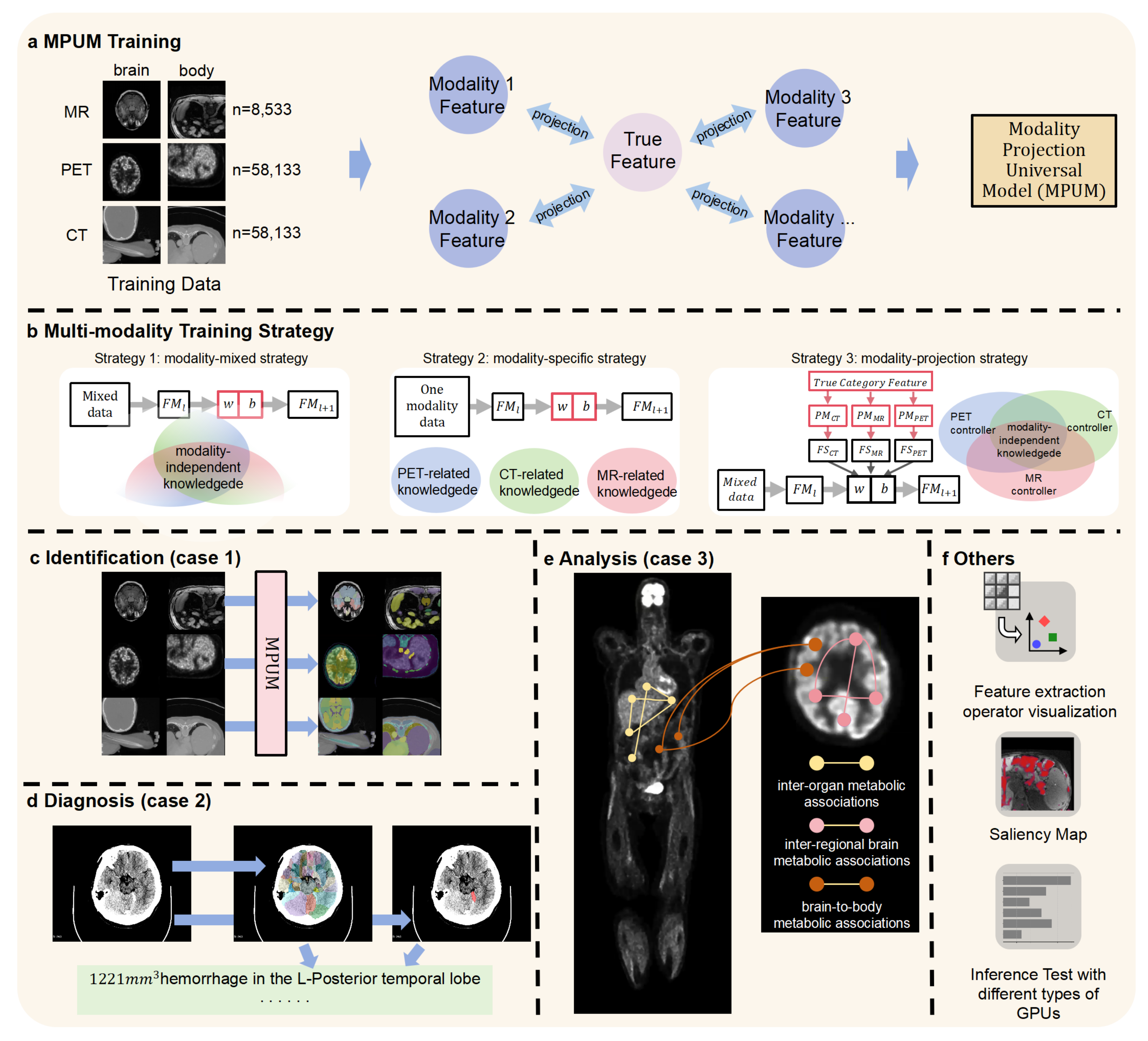}
    \caption{Overview of the modality projection universal model. \textbf{a,} Training process of the MPUM leveraging data from three distinct modalities. \textbf{b,} Comparison of two common multimodal data training strategies with our proposed modality-projection strategy. \textbf{c,} Application of the MPUM model as an aided identification tool across three modalities (over 500 categories). \textbf{d,} The MPUM model is utilized as an computer-aided diagnosis (CAD) tool for precise localization of intracranial hemorrhage with CT scans. \textbf{e,} Application of the MPUM Model as an aided analysis tool in identifying altered metabolic correlations in regions affected by epilepsy. \textbf{f,} Additional experimental results, including t-SNE visualizations of feature extraction operators and analysis of the network's saliency map. }
    \label{fig:overview}
\end{figure*}

With the efforts of the relevant communities, annotated medical multi-task datasets have become relatively mature. The TotalSeg dataset\cite{totalsegmentator} includes segmentation annotations for 104 types of CT anatomical structures from 1,204 unique subjects. The DAP dataset\cite{dap} covers 133 types anatomical structures with 533 CT scans. The CDUM dataset\cite{cdum} combines multiple single-task CT datasets to build a multi-task model capable of recognizing 25 organs and 6 types of tumors. The TotalSegMRI dataset\cite{totalsegmentatormri} covers 59 anatomical structures from 298 MR scans. Compared to single-task segmentation tasks, multi-task training offers greater robustness and performance\cite{bodymaps24,autopet,stunet,pcnet}. The performance boost of universal models from multi-task training primarily stems from two factors: firstly, the larger scale of multi-task datasets; secondly, the synergy between tasks, such as shared segmentation boundaries of adjacent tissues.

Current research has proposed seversal designs for universal medical models. CDUM\cite{cdum} integrates text embeddings into segmentation models to capture anatomical relationships. STUNet\cite{stunet} is a scalable and transferable U-Net model series, with sizes ranging from 14 million to 1.4 billion parameters. SAT\cite{sat} is designed to segment a wide array of medical images using text prompts. PCNet\cite{pcnet} utilizes prior category knowledge to guide the universal segmentation model in capturing inter-category relationships. Among the existing research on universal medical models, SAT is trained as a multi-task model on multi-modality data, whereas others are focused on single-modality data. Multi-modality datasets, being larger in scale compared to single-modality datasets, thus possess greater potential to achieve superior performance\cite{sat}. However, training with multi-modality data also presents greater challenges compared to single-modality training, such as differing feature distributions across modalities and increased training instability.

To address these limitations, we propose a versatile medical segmentation model based on a multi-modality projection mechanism. This mechanism allows for the extraction of modality-specific features from a shared high-dimensional space, enabling seamless generalization across different imaging modalities without the need for fine-tuning. Each organ has a high-dimensional latent feature, which can be projected in different directions to generate unique feature representations for each modality. This multi-modality projection allows for a unified understanding across different imaging techniques. Our modality projection universal model (MPUM) was trained using data from 861 unique subjects. Fig. \ref{fig:overview} illustrates the methodological innovations and experimental designs. Our proposed MPUM is designed with two key characteristics: precise brain segmentation and comprehensive whole-body segmentation. To demonstrate the clinical impact of MPUM, we focus on the following three key aspects: identification (Case 1), diagnosis (Case 2), and analysis (Case 3).

\begin{itemize}
    \item Case 1 (Fig. \ref{fig:overview}c) includes a technical validation of the segmentation performance of MPUM and other advanced universal models.
    \item Case 2 (Fig. \ref{fig:overview}d) demonstrates the process of MPUM aiding in identifying intracranial hemorrhage areas and brain region atlas based on CT scans.
    \item Case 3 (Fig. \ref{fig:overview}e) reveals the potential of MPUM as an aided analysis tool. Epilepsy is a systemic metabolic disorder \cite{lu2023glucose}, and whole-body analysis offers a more comprehensive understanding of its impact. We studied the impact of pediatric epilepsy on metabolism correlations through whole-body PET/CT scans. 
    
\end{itemize}

For ICH, our model addresses a critical challenge in the emergency room, where rapid diagnosis is vital, but radiologists may experience delays. MPUM enhances diagnostic efficiency by accurately identifying hemorrhages in CT scans, enabling quicker decision-making and timely intervention. Additionally, our team’s long-standing focus on the brain-body axis, which regulates multiple physiological systems, has led us to explore epilepsy. We specifically investigate metabolic changes outside the seizure focus, with MPUM revealing how these changes, including brain-body metabolic associations, correlate with epileptic activity. Overall, the MPUM model has transformative potential for clinical workflows, reducing manual annotation and improving diagnostic accuracy.


\section{Results}\label{sec:results}

We developed a deep-learning multi-modality universal segmentation model. In the identification tasks, the model demonstrates superior performance in anatomical structure identification,  outperforming existing segmentation models in terms of Dice and surface Dice metrics. In aided diagnosis, it accurately detected and quantified intracranial hemorrhages in CT scans, significantly improving diagnostic accuracy among general practitioners in a clinical setting. For aided analysis, the model facilitated the identification of significant alterations in metabolic associations in pediatric epilepsy, revealing metabolic or functional changes in connectivity of ROIs across the whole body. Furthermore, this section includes interpretable insights through visualization of feature operators and saliency maps.


\subsection{Identification: Anatomical Structure Identification (Case 1)}
\begin{figure*}[!t]
    \centering
    \includegraphics[width=\textwidth]{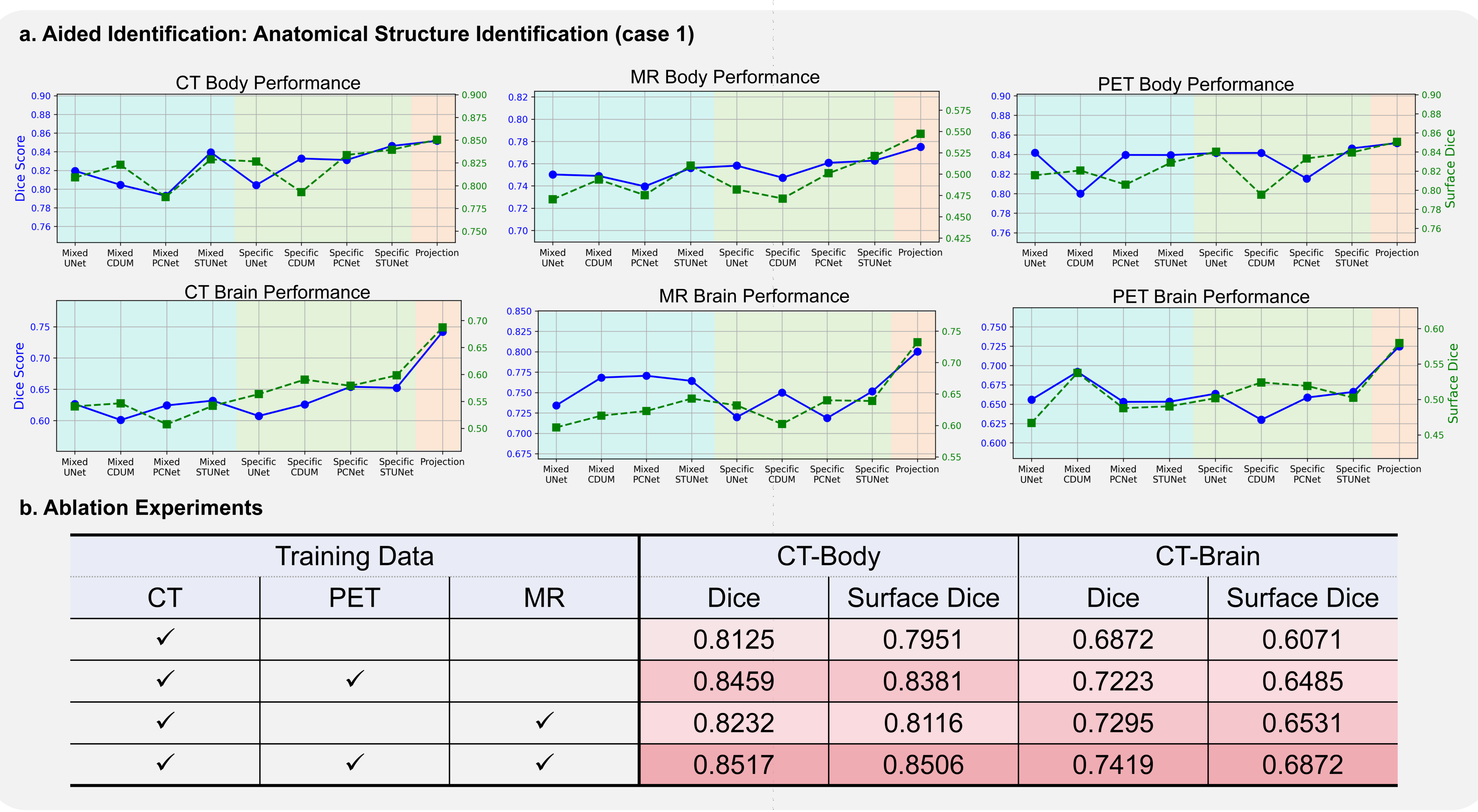}
    \caption{Comparison of the performance between the MPUM and other advanced models. \textbf{a,} Dice Score and surface Dice of  CT(left), MR(middle), and PET(right) imaging segmentation tasks. They compare four advanced models: UNet, CDUM, PCNet, and STUNet, with three multi-modality training strategies: mixed, specific, and projection. The modality-projection strategy represents the MPUM model. \textbf{b,} the table demonstrates the impact of using multi-modality training data on model performance.}
    \label{fig:case1}
\end{figure*}

We compared with state-of-the-art universal segmentation models, including CDUM\cite{cdum}, PCNET\cite{pcnet}, and STUNet\cite{stunet}, as well as the classic UNet\cite{unet} (with parameters adjusted to match the complexity of the other models). In addition, we evaluated different training strategies for multi-modality data: the modality-mixed strategy, the modality-specific strategy, and our proposed modality-projection strategy.

\textbf{Comparison between different models.} As shown in Fig. \ref{fig:case1}a, for MRI body segmentation, our projection strategy outperformed all other approaches, achieving the highest Dice score of 0.7751 and the highest surface Dice score of 0.5471. In comparison, the best-performing mixed strategy model, STUNet, achieved a Dice of 0.7560 and a surface Dice of 0.5100, while the best modality-specific strategy, STUNet, reached a Dice of 0.7627 and a surface Dice of 0.5211.

Similarly, in the CT body segmentation, the projection strategy again demonstrated superior performance, with a Dice score of 0.8517 and a surface Dice score of 0.8506. The closest competitor, the modality-specific STUNet, achieved a Dice of 0.8462 and a surface Dice of 0.8395, whereas the mixed strategy STUNet model recorded a Dice of 0.8394 and a surface Dice of 0.8288.

In the case of CT brain segmentation, the projection strategy continued to yield the highest scores, with a Dice of 0.7419 and a surface Dice of 0.6872. The highest scores among the other strategies were from the modality-specific PCNET, which achieved a Dice of 0.6540 and a surface Dice of 0.5982, and the mixed strategy STUNet, with a Dice of 0.6318 and a surface Dice of 0.5417. Overall, these results consistently demonstrate that our projection strategy outperforms both mixed and modality-specific strategies. 

We observed that, for multimodal data, using the mixed strategy with the same model tends to underperform compared to the specific strategy. For instance, in the MRI body segmentation, the Dice score for the mixed strategy using PCNET was 0.7396, whereas the specific strategy achieved a higher Dice of 0.7607. Similarly, in the CT body segmentation, the mixed strategy CDUM yielded a Dice score of 0.8046, while the specific strategy CDUM improved upon this with a Dice score of 0.8327. This performance decline in the mixed strategy may be attributed to the interference between different modality data distributions. It is difficult for the model parameters to reach an optimal state for all modalities simultaneously. For example, when the input is a CT image compared to when it is an MR image, there are significant differences in the optimization gradients for the same parameter, which affects the optimization process. In contrast, the projection strategy effectively mitigates this interference, proving to be a robust approach for training with multi-modality data.

\textbf{Ablation experiments. } As shown in Fig. \ref{fig:case1}b, the inclusion of multimodal training data significantly enhances the performance of image segmentation models. The use of CT alone yielded a Dice score of 0.8125 and a surface Dice of 0.7951 for CT-Body segmentation tasks, and a Dice score of 0.6872 and a surface Dice of 0.6071 for CT-Brain segmentation tasks. With the incorporation of PET data (CT\&PET), there is a noticeable improvement in both Dice and surface Dice scores for CT-Body (Dice: 0.8459, surface Dice: 0.8381) and CT-Brain (Dice: 0.7223, surface Dice: 0.6485). The integration of MR data with CT (CT\&MR) similarly enhances the model performance, particularly in CT-Brain (Dice: 0.7295, surface Dice: 0.6531) compared to CT-only model. The best performance is observed when all three modalities are used together (CT\&PET\&MR), achieving the highest Dice and surface Dice scores. These results suggest that multi-modality training could be particularly beneficial for enhancing segmentation tasks in medical imaging.



\subsection{Diagnosis: intracranial hemorrage (Case 2)}

\begin{figure*}[!t]
    \centering
    \includegraphics[width=\textwidth]{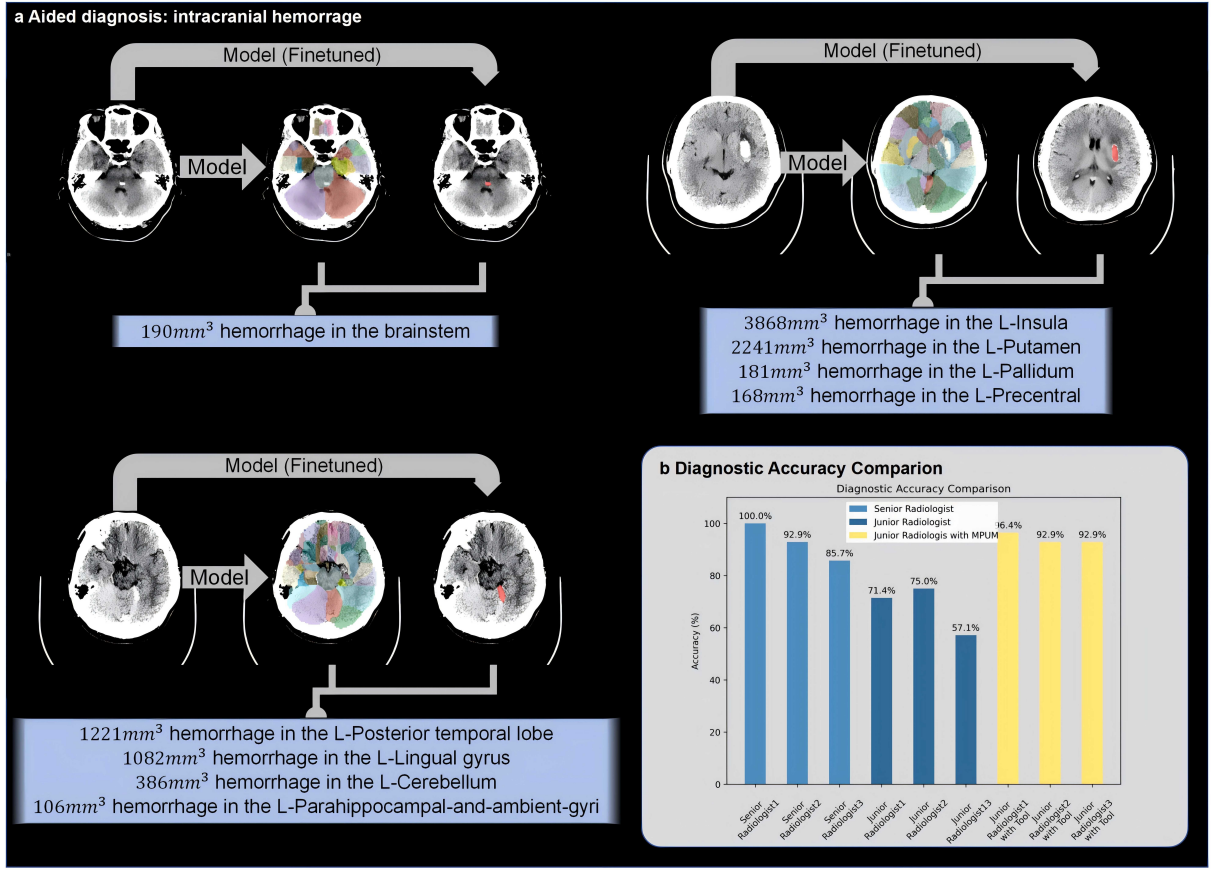}
    \caption{Performance of the MPUM Framework as aided diagnosis tool. \textbf{a}, Utilization of the MPUM as an aided diagnosis tool to detect hemorrhages and map brain regions from CT head scans, facilitating a precise diagnosis automatically. \textbf{b}, Illustration of the impact of the MPUM framework on enhancing diagnostic accuracy and support to general doctor in real-world settings.}
    \label{fig:case2}
\end{figure*}

Intracranial hemorrhage (ICH) is a life-threatening emergency where quick diagnosis and treatment decisions are critical for patient survival\cite{greenberg20222022,hemphill2015guidelines,sanner2024voxel}. ICH has a high mortality rate, with up to 40\% of patients dying within a year of the event\cite{an2017epidemiology,pinho2019intracerebral}, and a significant proportion of survivors suffering from lasting functional impairments\cite{moon2008prehospital}. Quick and accurate identification of ICH is essential for timely medical intervention, which can significantly improve patient outcomes\cite{hillal2022computed,sage2020intracranial,zimmerman2006radiologic}. The goal of Case 2 is to enable the MPUM model to provide doctors with an aided diagnosis under the objectives of speed and accuracy. The experimental setup for Case 2 consists of two parts: firstly, we tested 100 cases on the Instance2022 dataset\cite{instance2022_1,instance2022_2} and conducted statistical analyses on the volume of brain regions affected by ICH. Secondly, we validated the effectiveness of the MPUM with the validation of three senior radiologists and three junior radiologists on the in-house ICH dataset from emergency department.

\textbf{Precise hemorrhage measurement.} Our MPUM framework was fine-tuned on the Instance2022  dataset to identify hemorrhagic areas from CT head scans. As depicted in Fig. \ref{fig:case2}\textbf{a}, the original well-trained MPUM provides brain regions maps from CT scans, while the finetuned MPUM yields segmentation map for hemorrhages. By integrating these two predictions, we obtained precise aided diagnosis results. For instance, in the second scan of Fig. \ref{fig:case2}\textbf{a}, the model autonomously detected a hemorrhage involving 3868 $mm^3$ in the left insula and 2241 $mm^3$ in the left putamen. Precise quantification of ICH through CT scans plays a critical role in clinical decision-making and treatment planning\cite{sage2020intracranial}. Accurate volume measurements are essential for determining the extent of hemorrhage, monitoring its progression, and evaluating the risk of further complications like hematoma expansion, which is closely associated with worse outcomes\cite{hillal2022computed}. We conducted a detailed analysis of hemorrhage volumes across various brain regions in the Extended Data Fig.\ref{fig:extendedinstance2022}. Some regions, such as the insula, show relatively higher average volumes, which aligns with clinical observations where certain brain areas are more prone to larger hemorrhages due to their vascular structures and the prevalence of small vessel disease\cite{hillal2022computed,prakash2012segmentation}. These detailed quantification has proven crucial for developing more targeted therapies and enhancing diagnostic accuracy\cite{kim2021cerebral}.

\textbf{Enhancing diagnostic accuracy. } Additionally, we collected 28 ICH CT scans, along with diagnostic reports, from the emergency department of an external medical center. The diagnostic reports contain precise diagnostic results assisted by MRI imaging, which is regarded as ground truth. In the diagnostic accuracy test, three experienced radiologists with more than five years of experience and three general radiologists analyzed these cases, assessing the brain regions affected by ICH. Each physician should determine one or muptiple areas that ICH region involves. The regions that can be selected are: frontal lobe hemorrhage, temporal lobe hemorrhage, parietal lobe hemorrhage, occipital lobe hemorrhage, basal ganglia hemorrhage, cerebellar and brainstem hemorrhage, subarachnoid hemorrhage, subdural hemorrhage, and ventricular hemorrhage. Senior radiologists achieved accuracy rates of 100\%, 92.9\%, and 85.7\%, while the junior radiologists initially reached 71.4\%, 75.0\%, and 57.1\%. Most junior radiologists' errors were due to not fully identifying all hemorrhage regions and confusion between brain regions. In a specific case, while the frontal lobe hemorrhage and subarachnoid hemorrhage are identified, the ventricle hemorrhage may be overlooked.

With the aid of the MPUM aided diagnostic tool, the junior radiologists' accuracy improved to 96.4\% and 92.9\%, and 92.9\%, demonstrating the efficacy and potential of the MPUM framework in aided diagnosis. Additionally, the MPUM can identify brain regions with greater precision, distinguishing up to 83 different areas.

\subsection{Analysis: multi-organ metabolic associations (Case 3)}

\label{sec:auxiliaryanalysis}

\begin{figure*}[!t]
    \centering
    \includegraphics[width=\textwidth]{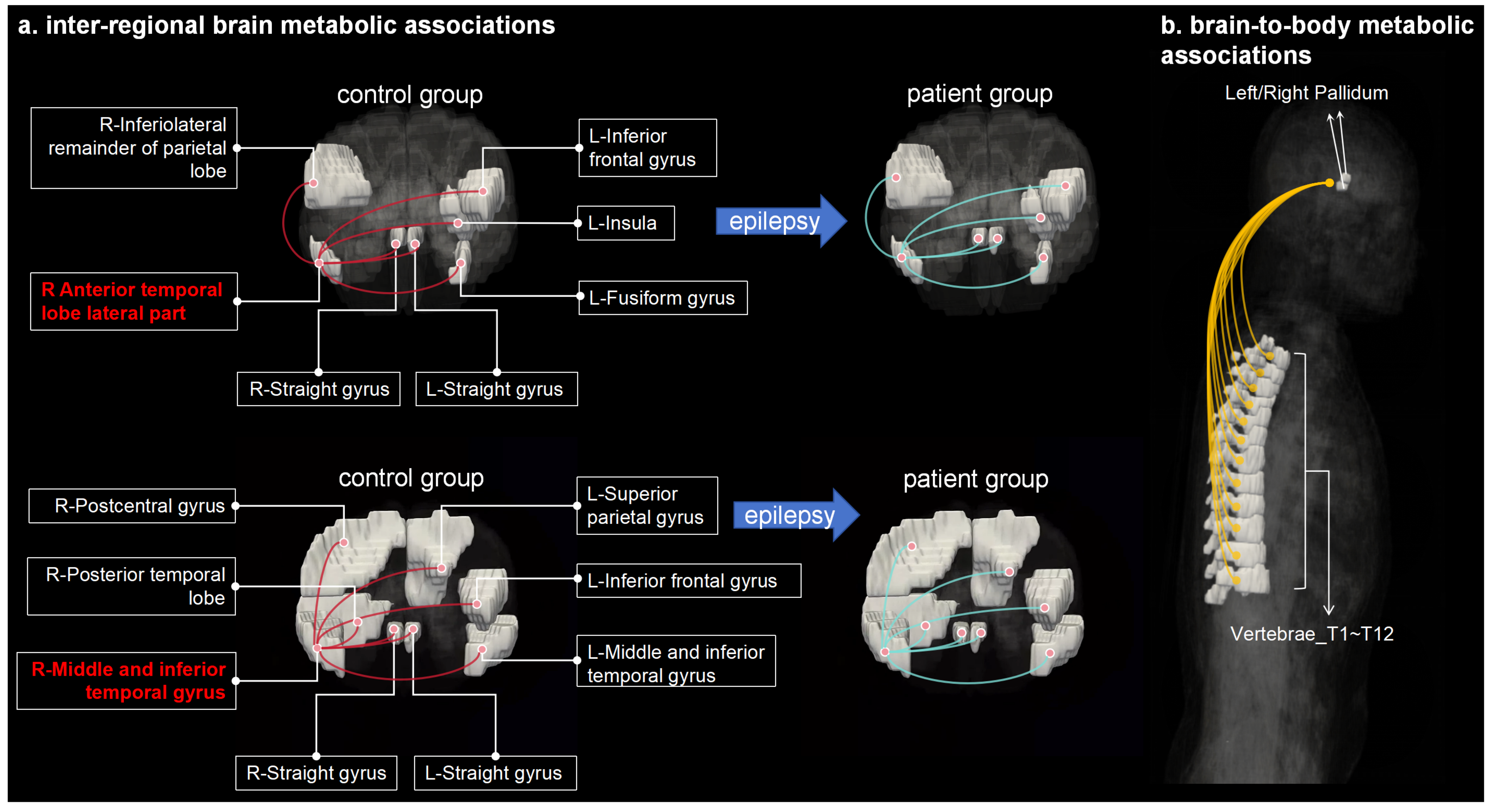}
    \caption{Multi-organ metabolic association analysis for pediatric epilepsy based on the universal model. We analyzed the metabolic associations in the epilepsy patient group (n=50) and the control group (n=22), using Fisher Z-Transformation to calculate the significance of differences in Pearson correlation coefficients. \textbf{a\&b,} Schematic representation of the connectivity among brain regions associated with the Right Anterior Temporal Lobe Lateral Part and the right middle and inferior temporal gyrus. The left diagrams illustrate the strong metabolic connection within the control group. Notably, these correlations are statistically significantly reduced in the patient group (p\textless0.001). \textbf{c,} Metabolic connectivity between the Pallidum and Vertebrae T1-T12 affected by epilepsy.}
    \label{fig:qianfoshan}
\end{figure*}


We assessed the efficacy of our universal model as an aided analysis tool in pediatric epilepsy research, specifically analyzing the impact of epilepsy on multi-organ metabolic associations. Utilizing PET and CT imaging for multi-organ metabolic analysis is vital due to its ability to assess multiple organs synchronously, enabling the exploration of systemic, inter-organ interactions\cite{sundar2023whole}. This technology is essential for understanding diseases that impact multiple body systems, like cancer and other systemic diseases. By using PET imaging data, researchers have established methods to study systemic metabolic abnormalities and inter-organ communication, which are crucial in understanding complex diseases\cite{sun2022identifying}.

Our universal model facilitates rapid identification of ROIs in human tissue structures, significantly reducing the manpower costs. We analyzed whole-body PET/CT data from a control group (n=33) and a group of patients without active epilepsy episodes (n=55). Utilizing our well-trained universal model, we identified 215 ROIs in the CT scans. Subsequent analysis using metabolic information from PET scans compared metabolic associations in these 215 ROIs between the control and patient groups.


In the analysis of 215 ROIs, which included 83 brain regions and 132 body regions, we excluded 12 ROIs due to high background radioactivity, including the bladder, inferior vena cava, aorta, and pulmonary artery, leaving 203 effective ROIs. We defined metabolic associations between two different regions as one pair, such as the left kidney and liver. In total, we analyzed 20,503 pairs for metabolic associations, which comprised 3,403 brain-brain pairs, 7,140 body-body pairs, and 9,960 brain-body pairs. Given that epilepsy is a brain-triggered disorder, our analysis focused on brain-related pairs, excluding the 7,140 body-body pairs.  To the best of our knowledge, this might be the largest effort for multi-organ metabolic analysis of epilepsy.

\textbf{The analysis of brain-brain pairs.} We examined whether the metabolic associations in each pair were significantly altered due to epilepsy. Among the 3,403 brain-brain pairs, 228 pairs showed significant changes in correlation due to epilepsy (p\textless0.001). Notably, 108 pairs involved the `right anterior temporal lobe lateral part', and 68 pairs involved the `right middle and inferior temporal gyrus' (see Extended Data Fig.\ref{fig:daixieguanlian}a-b). These findings suggest a strong link between the metabolic activity in these brain areas and epilepsy since epilepsy significantly alters the correlation. Fig. \ref{fig:qianfoshan}a illustrates the metabolic associations of the `right Anterior temporal lobe lateral part' with other brain regions. For the control group, there are high correlations between the `right anterior temporal lobe lateral part' and other brain regions, whereas the patient group showed weak metabolic connections. Fig. \ref{fig:qianfoshan}a also  displays a similar phenomenon centered around the `right middle and inferior temporal gyrus'. Previous studies indicate that both `right anterior temporal lobe lateral part' and `right Middle and inferior temporal gyrus' are common sites for epileptic foci\cite{roper1993surgical,engel1996introduction,engel1996surgery}. 


\textbf{The analysis of brain-body pairs. }We also analyzed metabolic associations between brain regions and body structures in 9,960 pairs. Interestingly, we identified significant changes in metabolic associations in 14 pairs (p\textless0.001). As illustrated in Fig. \ref{fig:qianfoshan}c, a notable metabolic association change was observed between the `pallidum' and `vertebrae T1 to T12'. This finding suggest a significant disruption in metabolic connectivity, potentially due to neuronal dysfunctions in epilepsy, where abnormal neuronal discharges might interfere with normal neural circuits. This disruption could affect signal transmission between the pallidum and the spinal region, thereby altering their metabolic association. These changes in metabolic connectivity, along with their correlation coefficients, are detailed in Extended Data Fig.\ref{fig:daixieguanlian}c. 

\textbf{The metabolic analysis of single organ.} Furthermore, we conducted a significant analysis of metabolic changes within individual organs between the control and patient groups, as detailed in Extended Data Fig.\ref{fig:danqiguandaixie}. We discovered that epilepsy not only causes metabolic abnormalities in certain brain regions but also leads to significant metabolic changes in other parts of the body, such as in some bones and muscles. Epilepsy increases the cereral metabolic rate of oxygen and ATP demand, leading to mitochondrial exhaustion, which might help substantiate changes in some brain regions, such as temporal lobe and thalamus\cite{liotta2024metabolic}. Additionally, metabolic dysfunctions also affect various bodily functions, including muscle activity\cite{fei2020metabolic}. 

\subsection{Interpretability}

\subsubsection{Visualization of feature operators}

The modality-projection model incorporates a controller module that transforms features from the modality space into feature extraction operators. The distribution of these feature operators across different modalities, as visualized on Fig. \ref{fig:saliency}, enhances the interpretability of the model.

Stage 1 represents the shallowest convolutional layer, where the feature extraction operators primarily extract low-level semantic information. Conversely, stage 4 denotes the deepest convolutional layer, tasked with processing high-level semantic information. The primary role of shallow convolutional layers is to extract low-level features from images, such as edges, textures, and simple shapes. These features are typically modality-independent, indicating a significant similarity in the low-level features extracted by different imaging modalities within these layers. For MR and PET modalities, there is an overlap in the feature operators, as both provide high contrast in imaging soft tissues. Their similarity in imaging characteristics of soft tissues might account for the observed overlap in their feature operators on the t-SNE plots. Although CT is better suited for depicting hard tissues, it extracts basic edges and texture features in the shallow layers, which may show partial similarity to those features extracted in the MR and PET modalities.

For deep convolutional layers (stage 3 and stage 4), these layers are tasked with extracting high-level features, which capture specific semantic information pertinent to each modality. The feature extraction operators in deep convolutional layers focus on modality-specific high-level features, leading to a pronounced separation of the feature operators for MR, PET, and CT in the t-SNE plots. Each modality's feature operators form distinct clusters at this level, reflecting the inherent differences in advanced feature extraction across modalities. 

This analysis of the feature operator highlights the contrast between the universality of low-level feature extraction and the specificity of high-level feature extraction. Furthermore, the interplay between this universality and specificity can be reconciled within the framework of a modality-projection controller.

\subsubsection{Saliency map}

\begin{figure*}[t]
    \centering
    \includegraphics[width=\textwidth]{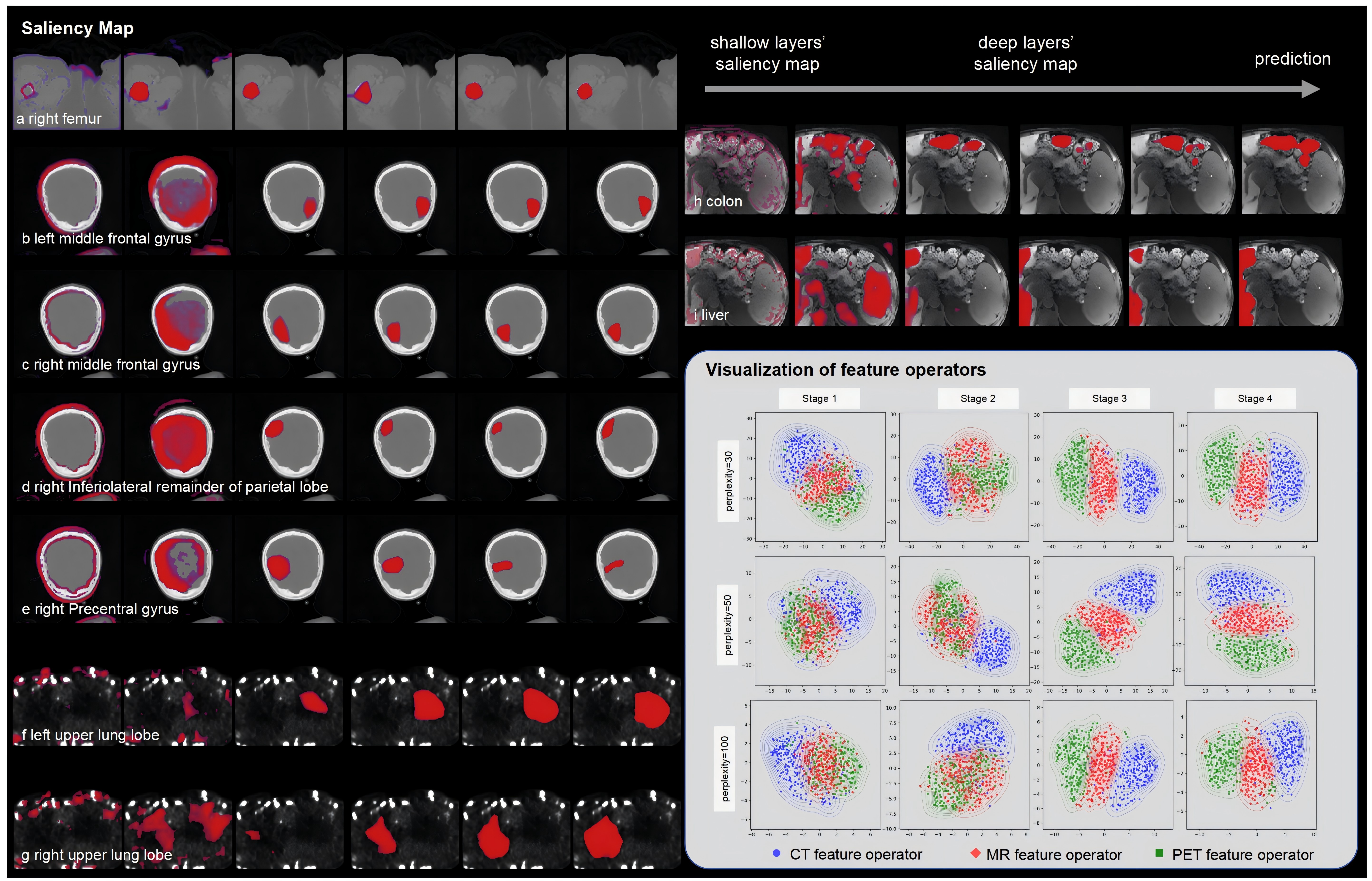} 
    \caption{Comprehensive visualization of saliency maps and feature operators. The black region displays a progression of saliency maps from shallow to deep layers, encompassing nine cases: \textbf{a,} a body CT scan, \textbf{b-e,} brain CT scans, \textbf{f\&g,} PET scans, \textbf{h\&i,} MR scans. The white region showcases the t-SNE visualization of convolutional kernel operators, capturing the feature extraction from shallow to deep layers.}
    \label{fig:saliency}
\end{figure*}


To illustrate the details of our model's reasoning and enhance the interpretability of our universal medical model, we present the saliency maps. Using the advantages of the modality projection controller, we can directly display the saliency maps of each neural network layer for specific categories. In contrast to traditional gradient-based CAM methods \cite{gradcam} that showcase the saliency of the final decision layer, our approach provides a stronger interpretability throughout the entire inference process of the model. As shown in Fig. \ref{fig:saliency}, we plot the saliency maps for different regions under various modalities. Our analysis revealed that saliency differences are related to layer depth, modality, and organ category.


In the shallow layers of the network, saliency is primarily focused on the image's contours and textures, while in the deeper layers, saliency increasingly concentrates on the ROI regions. As shown in Fig. \ref{fig:saliency}, the saliency map of the shallowest layer (the leftmost column) highlights the edge information of the image. The subsequent layer focuses on the texture information over large areas of the image (second column from the left). In contrast, the deeper network layers progressively refine the ROI regions with increasing precision. This progression indicates that the shallow layers capture global edge and texture information, whereas the deeper layers utilize local information to fine-tune the prediction results.


As shown in Fig. \ref{fig:saliency}a, the task of identifying the right femur from a CT image is relatively straightforward due to the high density of bone structures. In contrast, Fig. \ref{fig:saliency}b-e depict the identification of various brain regions, challenging tasks given the difficulty of distinguishing soft tissue areas from CT images. In Fig. \ref{fig:saliency}a, the shallow layers recognize a small amount of contour information, which is sufficient to clearly identify the location of the femur. Conversely, Fig. \ref{fig:saliency}b-e need recognize the overall contour of the skull for preliminary localization, followed by a focus on the subtle variations in the soft tissue across the entire brain region. 


Fig. \ref{fig:saliency}a-e represent cases from the CT modality, while Fig. \ref{fig:saliency}f-g are from the PET modality and Fig. \ref{fig:saliency}h-i. are from the MR modality. In the CT modality, distinct contour information results in saliency maps showing continuous red lines. The saliency maps of MR modality show the similar results. In contrast, the PET modality uses high metabolic activity points as reference, with shallow network layers focusing saliency around these points for initial localization.

\section{Discussion}\label{sec12}

Our study demonstrates that the proposed MPUM, employing a modality-projection training strategy, achieves robust performance. The experimental results highlight significant advantages of this strategy in multi-modality training. This model shows potential in medical imaging tasks\cite{cdum,pcnet,totalsegmentator}, extending the applications of  traditional universal segmentation models. The experimental results highlight two key advantages of this approach: precise brain segmentation and comprehensive whole-body segmentation. Specifically, the model enhances diagnostic accuracy for intracranial hemorrhage (ICH), supporting physicians in timely and accurate diagnosis. Additionally, it enables the analysis of systemic metabolic changes associated with epilepsy in pediatric patients, offering insights into the brain-body interactions in this context.


\subsection{Universal CT-based Brain Segmentation}
Previous studies have lacked efficient models for universal brain segmentation from CT scans, primarily due to the low contrast of soft tissues in CT images. Existing methods often rely on knowledge transfer, such as BraSEDA \cite{braseda}, which uses GANs to transfer MR brain region knowledge to CT, and UNSB \cite{unsb}, which transfers ventricle region knowledge via a diffusion Schrödinger bridge. However, these methods face two key challenges: the effectiveness of knowledge transfer depends on domain shifts, and they typically target only limited brain regions. MPUM addresses these issues, offering high accuracy and the ability to segment 83 distinct brain regions.

Additionally, accurate segmentation of brain ventricles in CT scans is critical for emergency procedures like ventriculostomy, used to treat conditions such as hydrocephalus, brain injury, and tumors \cite{unsb}. Unlike MRI, which is slower and less accessible, CT scans are faster and more readily available, making them crucial for rapid ICH diagnosis in emergency settings. Although MRI remains the gold standard for ICH detection, MPUM enables precise ventricle segmentation from CT scans, including the third ventricle, lateral ventricles, and temporal horn. This capability offers a significant clinical advantage, as quick and accurate diagnosis is essential for timely intervention and improved outcomes in emergency care \cite{hillal2022computed,sage2020intracranial,zimmerman2006radiologic}.


Furthermore, the key factors in predicting outcomes and treatment strategies for ICH are midline shift and hemorrhage volume \cite{midline1}. Specifically, hemorrhage volume in the brainstem is crucial in determining whether conservative medical treatment or surgical intervention is required \cite{midline1}. As shown in Case 3, MPUM is capable of assessing hemorrhage volume in brainstem, providing valuable information for clinical decision-making. While MPUM does not directly detect the midline, it can infer it by identifying left and right brain regions, such as the left and right cerebellum. The interface between these regions indirectly marks the brain’s midline.

\subsection{From Brain Imaging Analysis to Brain-body Axis Exploration}
MPUM excels in whole-body segmentation, which supports our team’s long-standing research on the brain-body axis—the bidirectional communication between the brain and body that underpins many physiological and psychological processes \cite{li2023mechanism,sharp2021connecting}. While epilepsy has traditionally been viewed as a brain-centric disorder, recent studies suggest it may also involve systemic metabolic effects beyond the brain \cite{lu2023glucose}. This insight prompted us to explore whole-body metabolic changes in epilepsy.
Leveraging MPUM’s ability to rapidly identify regions of interest (ROIs), we analyzed metabolic correlations between brain-brain and brain-body pairs in epilepsy patients. Notably, significant metabolic changes were observed in brain regions such as the right anterior temporal lobe and right middle/inferior temporal gyrus, which are consistent with existing research on epileptic foci \cite{roper1993surgical,engel1996introduction,engel1996surgery}. These findings enhance the role of imaging in epilepsy surgery planning, which typically integrates data from electroencephalography and clinical evaluations.

In addition to brain-brain correlations, we discovered significant metabolic changes in brain-body pairs, particularly between the pallidum and vertebrae T1–T12. Epilepsy patients are at greater risk of vitamin D deficiency, which impacts bone health, a risk exacerbated by antiepileptic medications that contribute to bone density loss \cite{sourbron2024vitamin}. To our knowledge, this connection has not been explored previously, highlighting the potential of a systematic whole-body metabolic analysis. This data-driven discovery opens new avenues for understanding epilepsy and identifying novel biomarkers for diagnosis.

\subsection{Limitations}
Despite the progress our model has made in clinical applications and medical analysis, several challenges remain. The integration of AI tools into existing medical workflows is complex. The model must not only be accurate but also user-friendly, providing outputs in a manner that seamlessly integrates with the workflows of healthcare professionals. Besides, as the model scales to handle more tasks, the computational demands increase, posing challenges for efficient operation on clinical hardware.

\subsection{Future}
In our future work, we plan to continue expanding multi-task learning. Currently, our universe model recognizes human tissues and organs. Furthermore, we intend to develop a framework for continual training to optimize the model's performance and stability consistently. Additionally, we will continue to explore the potential of the universal model as an aided analysis tool for downstream tasks. We believe that ongoing enhancements to the universe model will significantly improve its effectiveness and practicality in medical imaging research.

\section{Methods}\label{sec:method}
\subsection{Study population}
In Case 1 study, we employed the 18F-FDG PET/CT dataset\cite{autopet} alongside two MR datasets\cite{totalsegmentatormri,mribrain}. Specifically, we utilized 533 negative control subjects from the 18F-FDG PET/CT dataset\cite{autopet}. Each patient scan in $^{18}$F PET/CT dataset includes both whole-body CT and PET scans. The body region labels were sourced from DAP Atlas Dataset\cite{dap}, covering 142 distinct anatomical structures. DAP label dataset only contains the annotations of 533 subjects' scans. After excluding irrelevant labels such as ``background" and ``left to annotations", we utilized 133 effective labels from the DAP dataset. Given the DAP’s focus on body regions, it overlooked detailed brain regions. To address this gap, we pre-segmented the brain regions in the PET scans using the MOOSE tool\cite{moose,nnunet}, followed by refinement and correction by two experienced radiologists, resulting in 83 annotated brain regions. Consequently, the ``brain" label from the DAP dataset was removed, finalizing 132 non-brain categories and 83 brain region categories. The MR dataset includes 298 MR body scans\cite{totalsegmentatormri}, covering 43 body regions, and 30 MR brain scans, annotated with the same  83 brain regions\cite{mribrain}.
We utilized the Instance2022 \cite{instance2022_1,instance2022_2} dataset for our study (Case 2), which includes 100 non-contrast head CT volumes from clinically diagnosed patients with various types of intracranial hemorrhage (ICH), such as subdural, epidural, intraventricular, intraparenchymal, and subarachnoid hemorrhage. These CT volumes were sourced from Peking University Shougang Hospital, China, and meticulously labeled by 10 radiologists with over five years of clinical experience. We finetuned our MPUM model using the Instance2022 data, enabling it to accurately identify ICH. In addition, we obtained 28 ICH cases along with diagnostic reports from the emergency department of Peking University Third Hospital for further validation. The diagnostic reports contain precise diagnostic results assisted by MRI imaging.



The Case 3 of this study used 50 pediatric epilepsy patients recruited from Qianfoshan Hospital, Shandong, China, between August 3, 2021, and June 4, 2024. The cohort consisted of 27 females and 23 males, with ages ranging from 2 to 18 years ($11.78\pm5.94$ years). The height of the participants ranged from 0.93 meters to 1.85 meters ($1.49\pm0.236$ meters), and their weight ranged from 14.5 kg to 90 kg ($46.22\pm19.92$ kg). The control group comprised 22 participants, including 17 males and 5 females, with data collected from August 18, 2020, to May 9, 2024.

The multiple datasets used in this study, along with their descriptions, the division of training and testing data, and download URLs, are detailed in the Extended Data Fig.\ref{fig:extenddataset}.

\subsection{Ethical Approval Declaration}
This study was approved by the Ethics Committee of The First Affiliated Hospital of Shandong First Medical University \& Shandong Provincial Qianfoshan Hospital, with approval number S917. All participants provided informed consent, and the study adhered to all applicable ethical guidelines and regulations.

\subsection{Implements}
\subsubsection{Data preprocessing}
For this study, all imaging modalities, including CT, PET, and MR scans, underwent linear interpolation to achieve isotropic voxel sizes with a 2mm resolution. This standardization was essential to address variations in slice thickness and in-plane resolutions across different studies\cite{foundation_biomarker}. From these scans, we extracted patches of $128\times128\times128$ voxels, which correspond to a physical volume of $25.6 \times 25.6\times 25.6 \text{cm}^3$.

To ensure consistent and stable model training, we normalized the voxel values within these patches. For the CT patches, the voxel values were normalized between 0 and 1. For MR patches, the voxel values were divided by 3000 for normalization, while PET data were first converted to Standardized Uptake Value (SUV) and then divided by 20 for normalization. These normalization steps were crucial for maintaining data consistency across the different imaging modalities.

Additionally, we employed data augmentation techniques, specifically RandGaussianSmooth and RandAdjustContrast, to enhance the diversity and robustness of our training dataset. The RandGaussianSmooth method involves applying Gaussian smoothing to the images with a standard deviation randomly chosen between 0.5 and 1.5. This technique helps in reducing noise and simulating various levels of blurriness. The RandAdjustContrast method adjusts the contrast of the images by randomly scaling the intensity values between 0.5 and 1.5. These techniques partially simulate the effects of different imaging equipment and reconstruction algorithms, further enhancing the model's robustness.

We pre-cropped the CT, PET, and MR scans into standardized patches to improve training speed. Reading a $128\times128\times128$ patch is significantly faster than reading the entire image and then cropping it. This pre-processing step dramatically reduces the I/O time during training, allowing for more efficient use of computational resources. We have made the patch-wise multi-modality training dataset publicly available to facilitate further research and development in this area.

\subsubsection{Metrics}
We employed two evaluation metrics: Dice and surface Dice. Dice measures the overlap between predicted and true segmentations. It is calculated as twice the area of overlap between the two segmentations divided by the total number of pixels in both segmentations, providing an overall accuracy of how well the two align. Surface Dice, on the other hand, is a more specific measure that focuses on the boundary accuracy of the segmentation. It assesses how closely the boundaries of the predicted segmentation conform to the true surface contours of the object being analyzed.

\subsection{Modality Projection Principle}
\textbf{Motivation.} In recent studies of universal models, such as SAT \cite{sat} and CDUM\cite{cdum}, the modality-mixed strategy is commonly employed during the training phase. This strategy mixes data from various modalities, enabling a single model to adapt to data from multiple modalities. As shown in Fig. \ref{fig:overview}b, the modality-mixed strategy results in a multi-modality universal model. Although the modality-mixed strategy benefits from a larger training data set, it also encounter challenges related to feature interference among modalities. The feature extraction operators (weights and biases) must deal with imaging data from all three modalities. This may lead the model to prioritize modality-independent knowledge, potentially at the expense of modality-specific knowledge. In contrast, the modality-specific strategy involves training separate models for each modality. The performance of these models serves as the baseline for this study. As illustrated in Fig. \ref{fig:overview}b, there are three types of modality data—MR, PET, and CT—the modality-specific strategy could produce three modality-specific models: a PET model, a CT model, and an MR model. The drawbacks of training individual models for each modality is the limited amount of training data and the lack of feature collaboration between the multi-modality data. In the realm of multi-modality medical imaging, different imaging modalities such as CT, PET, MR, offer unique perspectives of the same underlying biological tissues. Each modality captures specific aspects of tissue characteristics due to differences in imaging principles and physical interactions. To effectively integrate information across modalities, we propose a modality projection theory centered on the concept of high-dimensional latent features that represent the comprehensive properties of each tissue type.

\textbf{Fundamental Principles.} The core of our theory is the assumption that for each tissue, there exists a high-dimensional latent feature vector $\mathbf{T} \in \mathbb{R}^{d_T}$, where $d_T$ denotes the dimensionality of the latent feature space. This latent feature encapsulates all intrinsic properties of the tissue that could be captured across various imaging modalities.

Each imaging modality provides a modality-specific projection of this latent feature into its own feature space. For a given modality $m \in {\text{CT}, \text{MR}, \text{PET}}$, we define a projection matrix $\mathbf{P}_m \in \mathbb{R}^{d_T \times d_m}$, where $d_m$ is the dimensionality of the modality feature space. The modality-specific feature vector $\mathbf{M}_m \in \mathbb{R}^{d_m}$ is obtained by projecting the latent feature:

\begin{equation} \mathbf{M}_m = \mathbf{T} \cdot \mathbf{P}_m. \end{equation}

This projection models how each modality perceives the underlying latent features of the tissue, capturing modality-specific characteristics.

\textbf{Inverse Projection and Reconstruction.} The relationship between latent features and modality-specific features suggests that, under certain conditions, it is possible to reconstruct the latent features from the modality-specific features using the inverse of the projection matrices:

\begin{equation} \mathbf{T} = \mathbf{M}_m \cdot \mathbf{P}_m^{-1}. \end{equation}

However, since each modality provides only a partial view, reconstructing $\mathbf{T}$ from a single modality may not capture all aspects of the latent features. Combining information from multiple modalities enhances the reconstruction, providing a more comprehensive representation.

\textbf{Extended Projection with External Models.} Optimizing $\mathbf{T}$ and $\mathbf{P}_m$ simultaneously can lead to instability, as the condition number of the system matrix may increase due to parameter interdependence. Changes in $\mathbf{T}$ propagate to $\mathbf{M}_m$, impacting the eigenvalues of $\mathbf{P}_m$, which amplify convergence difficulty and reduce gradient stability.

To address this challenge, we extend the projection model by incorporating features derived from external pre-trained models as additional projections of the latent features. Models like CLIP\cite{clip} and BioCLIP\cite{bioclip} provide robust feature embeddings for ROIs, trained on extensive datasets.

Let $\mathbf{M}_i \in \mathbb{R}^{d_i}$ represent the sub-space features from an external model $i$, with $d_i$ being its feature dimensionality. Each external model has its projection matrix $\mathbf{P}_i \in \mathbb{R}^{d_T \times d_i}$:

\begin{equation} \mathbf{M}_i = \mathbf{T} \cdot \mathbf{P}_i. \end{equation}

By leveraging these external features, we estimate the latent features by aggregating inversely projected sub-space features:

\begin{equation} \mathbf{T} = \frac{1}{N} \sum_{i=1}^{N} \mathbf{M}_i \cdot \mathbf{P}_i^{-1}, \end{equation}

\noindent where $N$ is the number of external models used. This approach anchors the latent features to stable, pre-trained representations, mitigating optimization instability. Additionally, the training datasets used by external pre-training models encompass a wide range of textual data, which facilitates the reconstruction of latent features.

\textbf{Generalized Modality Projection Framework.} In this generalized framework, the latent features $\mathbf{T}$ act as a central hub connected to various modality features $\mathbf{M}_m$ and external model features $\mathbf{M}_i$ through their respective projection matrices. This unified representation consolidates information across different modalities and knowledge domains.

\subsection{Modality Projection Universal Model Structure}
Building upon the modality projection principle, we introduce the Modality Projection Universal Model (MPUM), a comprehensive framework designed to effectively process and integrate multi-modality medical imaging data. The MPUM leverages the Modality Projection Controller (MPUM) within a deep learning architecture to handle diverse imaging modalities—such as CT, MR, and PET—facilitating accurate and efficient image segmentation.

As illustrated in Extended Data Fig.\ref{fig:extendstructure}a, the MPUM architecture begins with multi-modality patch inputs (e.g., CT, MR, PET), processed initially through a Head Layer, followed by a series of Dual-Branch Blocks with skip connections, and concluding with a Tail Layer to produce the segmentation output. The use of skip connections allows the model to preserve spatial information and integrate features from different layers, improving segmentation performance.

At the core of the MPUM is the Modality Projection Controller (MPC), illustrated in Extended Data Fig.\ref{fig:extendstructure}b. The MPC is responsible for dynamically adjusting the feature extraction process based on the specific characteristics of the input modality. 

The MPC process involves the following two steps. Firstly, the latent feature vector $\mathbf{T}$ is projected into the modality category space using the modality-specific projection matrix $\mathbf{P}_m$:

\begin{equation} \mathbf{M}_m = \mathbf{T} \cdot \mathbf{P}_m, \end{equation}
where $\mathbf{M}_m$ represents the modality-specific feature vector. Secondly, the modality features $\mathbf{M}_m$ are processed through a Multi-Layer Perceptron (MLP), serving as the Feature Operator Generator (FOG):

\begin{equation} 
\label{equ:6}
\mathbf{K}_m = \text{FOG}(\mathbf{M}_m), \end{equation}
resulting in convolutional kernels $\mathbf{K}_m$ tailored for each modality. This dynamic adjustment allows the MPUM to adapt its convolutional operations to the unique properties of each modality, thus improving feature extraction and segmentation accuracy.

The MPUM incorporates a specialized Dual-Branch Block Structure, illustrated in Extended Data Fig.\ref{fig:extendstructure}c, which consists of two parallel branches: traditional convolutional branch and controller-based convolutional branch. Traditional convolutional branch employs standard convolutional layers with fixed kernels to extract general features from the input data. It captures modality-invariant features. Controller-based convolutional branch utilizes the convolutional kernels generated by the MPC. The controller-based convolutional layers are non-parametric and adaptively adjust to the input modality, capturing modality-specific characteristics.

\subsection{Saliency map}

Saliency maps are vital tools for visualizing the decision-making process of deep learning models. Traditional methods for generating saliency maps primarily rely on gradient-based techniques. Methods such as Vanilla Gradients\cite{VanillaGradients} and Grad-CAM\cite{gradcam} generate saliency maps by calculating the gradients of the input image with respect to the model's output, highlighting regions that significantly influence the model's decisions.

As shown in Extended Data Fig.~\ref{fig:extendstructure}c, our modality projection block structure enables the generation of saliency maps across all layers of the network, not only the decision layer. In our model, we obtain the feature operator $\bm{K}$ by Equ. \ref{equ:6}. In our model, the feature operator $\bm{K}$ from Equ.\ref{equ:6} is a tensor with dimensions $\mathbb{R}^{C \times H \times 3 \times 3 \times 3}$, where $C$ represents categories, $H$ channels, and a kernel size of $3 \times 3 \times 3$.


\subsection{Statistical Analysis}
In Section \ref{sec:auxiliaryanalysis}, we applied the Fisher Z-transformation to compare correlation coefficients between two independent groups. This transformation standardizes raw correlation coefficients into values approximating a normal distribution, enabling hypothesis testing between control and patient groups.


In our analysis, we analyzed metabolic data from from organs $a$ and $b$ in a control group (33 individuals), calculating the mean standardized uptake value (SUV) for the regions of interest (ROI) within these organs. The metabolic data for organ $a$ in the control group is denoted as $A^{control} = \{a^n | n = 1, 2, \ldots, 33\}$, representing the set of mean SUVs for each individual. Similarly, the metabolic data for organ $b$ in the control group is represented as $B^{control} = \{b^n | n = 1, 2, \ldots, 33\}$. We calculated the Pearson correlation coefficient, which measures the strength and direction of a linear relationship between two variables:

\begin{equation}
\label{equ:pearson}
    r^{control}_{a,b}=\frac{\sum^n_{i=1}{(a^i-\overline{a})(b^i-\overline{b})}}{\sqrt{\sum^n_{i=1}{(a^i-\overline{a})^2(b^i-\overline{b})^2}}},
\end{equation}
where $a^i$ and $b^i$ are the individual SUV for organs $a$ and $b$ in the control group, and $\overline{a}$ and $\overline{b}$ are the mean SUVs for organ $a$ and $b$ respectively. We can simplify the Equ. \ref{equ:pearson} to $r^{control}_{a,b}=pearson(A^{control},B^{control})$. For the patient group (n=55), we can calculated the Pearson correlation coefficient $r^{patient}_{a,b}=pearson(A^{patient},B^{patient})$. 

We apply the Fisher Z-transformation to $r^{control}_{a,b}$ and $r^{patient}_{a,b}$, which converts the correlation coefficient to values that are normally distributed:

\begin{align}
    &z^{control}_{a,b} = arctanh(r^{control}_{a,b})\\
    &z^{patient}_{a,b} = arctanh(r^{patient}_{a,b})
\end{align}

We then calculated the Z-score, which is the standardized difference between the two Fisher-transformed correlation coefficients:

\begin{equation}
    z = \frac{z^{control}_{a,b}-z^{patient}_{a,b}}{\sqrt{{1}/{(n^{control}-3)}+{1}/{(n^{patient}-3)}}},
\end{equation}
where $n^{control}$ and $n^{patient}$ are the sample sizes of control group and patient group, respectively. Finally, we calculate the p-value for the Z-score under the normal distribution assumption:

\begin{equation}
    p = 2*(1-cdf(abs(z))),
\end{equation}
where $abs(z)$ represent the absolute value of the Z-score and $cdf$ stands for cumulative distribution function. The term $1 - cdf(abs(z))$ represents the one-tailed p-value, and multiplying by 2 makes it a two-tailed test.


\section{Acknowledgements}

Zhaoheng Xie discloses support for the research of this work from the Natural Science Foundation of China (62394311, 62394310),  Beijing Natural Science Foundation (Z210008), National Biomedical Imaging Facility Grant and from the startup funds of Peking University Health Science Center. Rui Wang discloses support for the research of this work from the NSFC Incubation Project of Guangdong Provincial People’s Hospital (KY0120220046) and from the Guangdong Basic and Applied Basic Research Foundation (2022A1515110674).

\section{Code availability}
All code used for data processing and performance analysis, including a well-trained model weights, is publicly available via GitHub at https://github.com/YixinChen-AI/MPUM under the MIT licence.

\nocite{*}
\bibliography{sn-mathphys-num}

\begin{appendices}




\clearpage
\begin{figure*}[t]
    \centering
    \includegraphics[width=\textwidth]{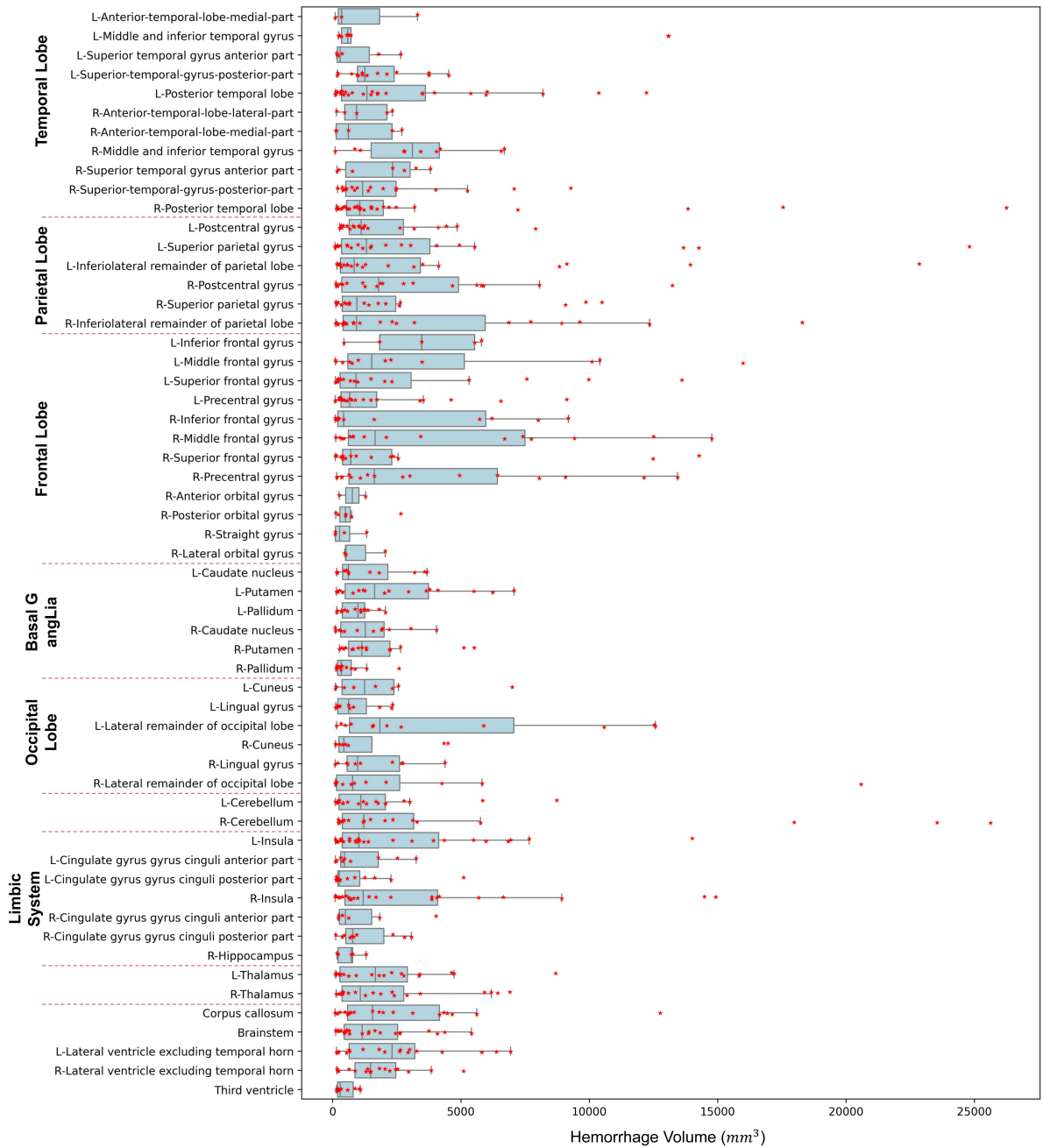}
    \caption{Boxplot of the brain region volumes where the ICH areas are located, organized by brain lobes. The figure summarizes data from 100 ICH cases included in the Instance2022 dataset. Each bar represents the average hemorrhage volume for that specific brain region, with the dots indicating individual measurements for each case.}
    \label{fig:extendedinstance2022}
\end{figure*}
\clearpage

\begin{figure*}[t]
    \centering
    \includegraphics[width=\textwidth]{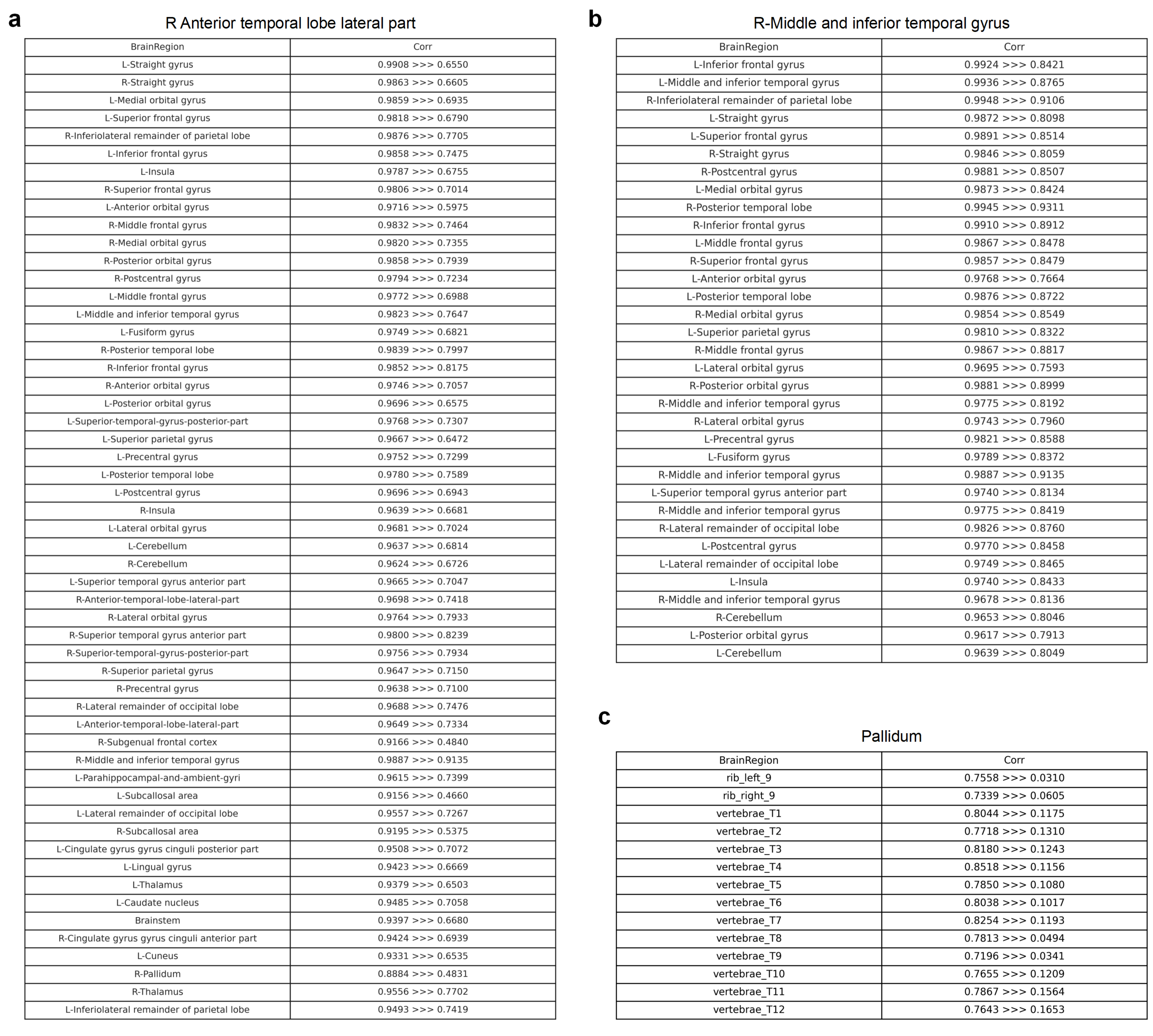}
    \caption{The impact of epilepsy on the metabolic correlations of organs. All results shown are statistically significant. The ``Corr" colume in the table is formatted as ``A>>>B," where value A represents the metabolic correlation between two regions under the control group, and value B indicates that in the patient group. \textbf{a,} displays all brain regions related to the ``R Anterior temporal lobe lateral part". \textbf{b,} displays all brain regions related to the ``right middel and inferior temporal gyrus". \textbf{a\&b} are the results of inter-regional brain metabolic associations (brain-brain pairs). \textbf{c,} displays significant change in brain-to-body metabolic associations (brain-body pairs).}
    \label{fig:daixieguanlian}
\end{figure*}
\clearpage

\begin{figure*}[t]
    \centering
    \includegraphics[width=\textwidth]{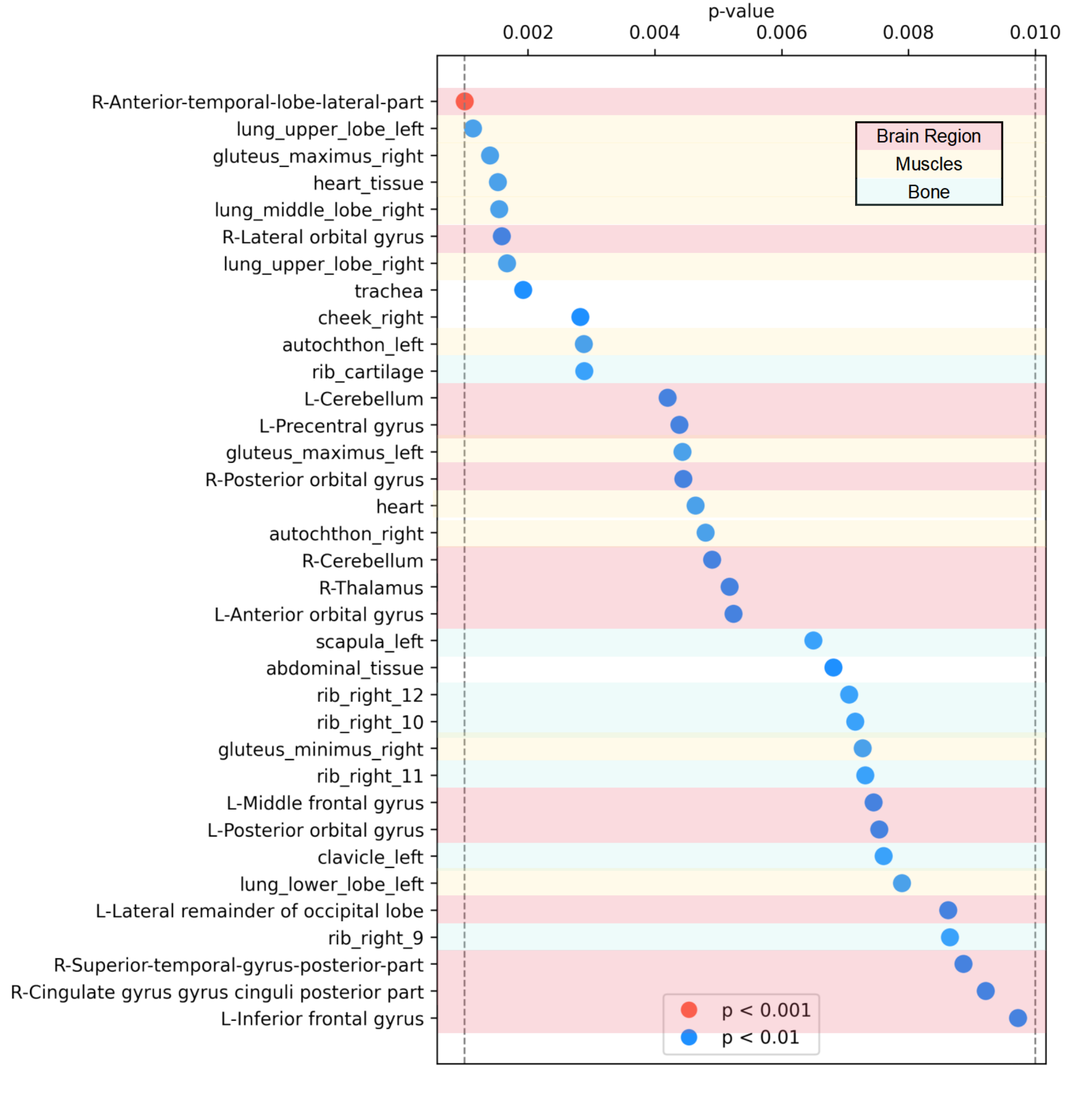}
    \caption{Significant metabolic changes in a single organ caused by epilepsy. Among a total of 215 ROIs, the figure displays the organs with statistically significant changes between control group and patient group. Each dot represents a different region, colored differently to indicate significance levels—red for p\textless0.001 and blue for p\textless0.01.}
    \label{fig:danqiguandaixie}
\end{figure*}

\clearpage
\begin{figure*}[t]
    \centering
    \includegraphics[width=\textwidth]{./extendeddata4.pdf}
    \caption{List of categories that the model can identify. The 215 categories includes 132 non-brain region categories and 83 brain region categories.}
    \label{fig:extendedcategories}
\end{figure*}
\clearpage
\begin{figure*}[t!]
    \centering
    \includegraphics[width=\textwidth]{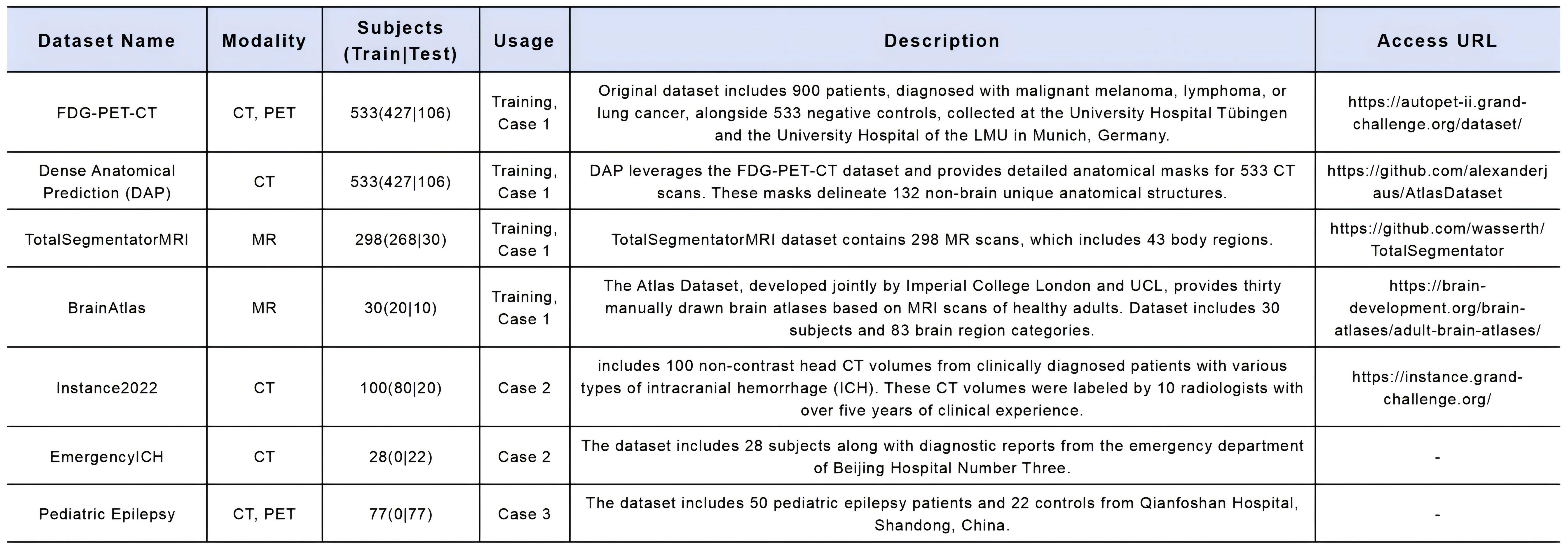}
    \caption{Detailed list of data used for training the model and in the three cases. The first four datasets were used in training the MPUM model, enabling it to recognize 215 categories. The Instance2022 trainset was used to fine-tune the MPUM model to specifically recognize ICH categories.}
    \label{fig:extenddataset}
\end{figure*}

\clearpage

\begin{figure*}[t]
    \centering
    \includegraphics[width=\textwidth]{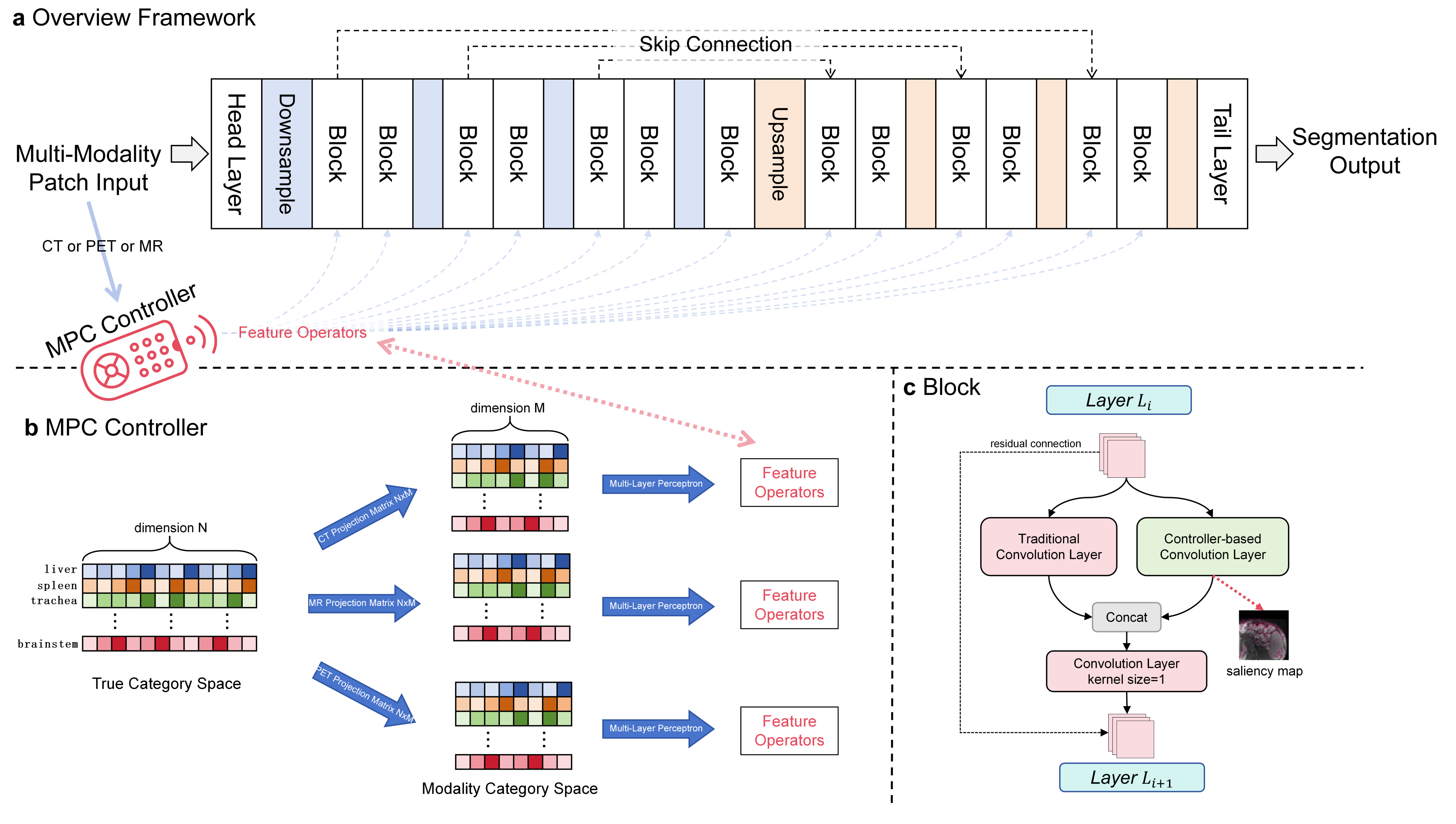}
    \caption{The detailed structure of the MPUM model. \textbf{a,} shows the overview framework of MPUM. The modality-projection controller influences the feature operators within each block based on the input modality. \textbf{b,} provides a detailed explanation of the controller mechanism. Each category has a latent feature with dimension N. Through different modality projection matrices, various modality feature spaces with dimension M can be obtained. Subsequently, these features are transformed into feature operators via a multi-layer perceptron. \textbf{c,} displays the detailed structure of the block. Each block contains two branches: one is a traditional convolutional layer, and the other is a non-parametric controller-based convolution layer, whose kernel is generated by the controller. The output feature maps from the two branches are merged through concatenation. The output feature map of the controller-based convolution layer also serves as the saliency map, providing interpretability.}
    \label{fig:extendstructure}
\end{figure*}

\clearpage

\begin{figure*}[t]
    \centering
    \includegraphics[width=\textwidth]{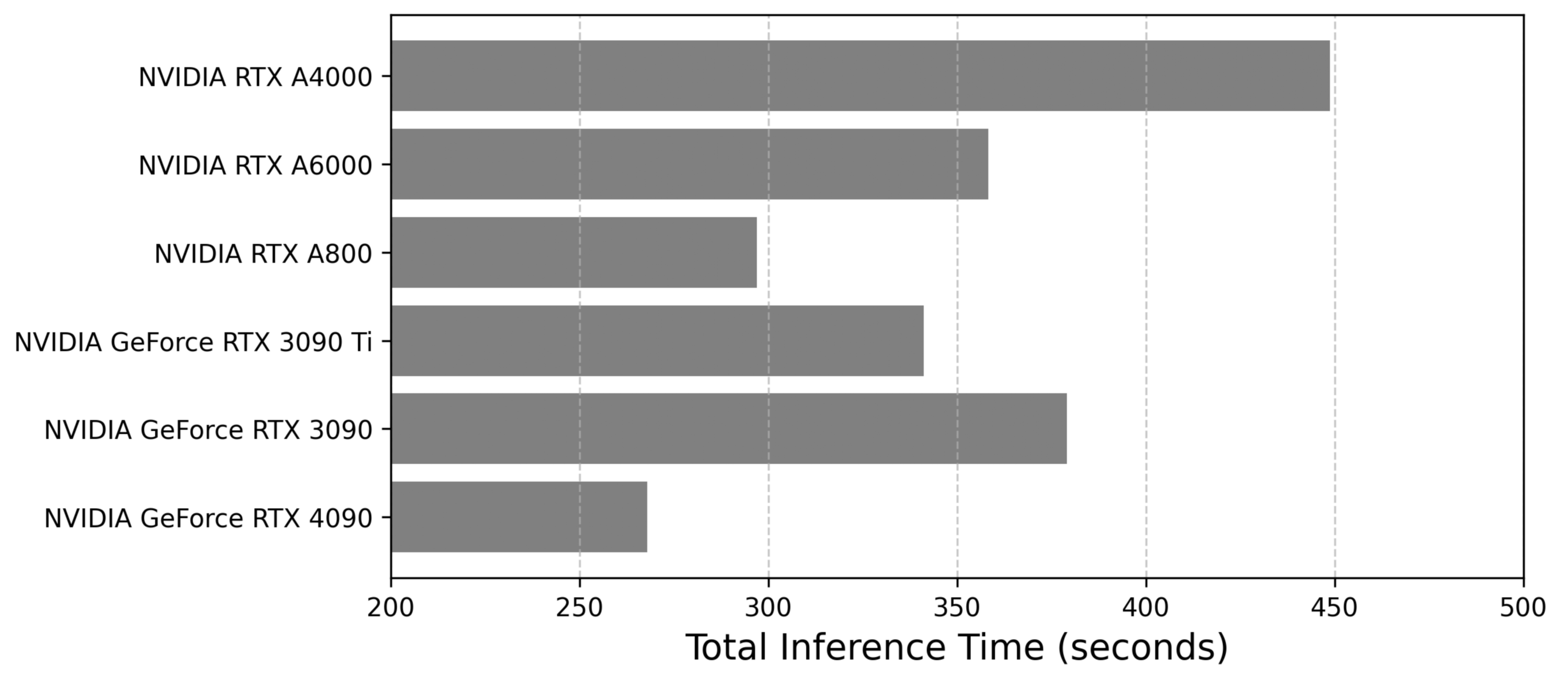}
    \caption{The inference time of MPUM when deployed on different graphics cards. MPUM requires 16GB of VRAM for deployment. The inference tests utilize 106 whole-body CT cases from the $^{18}$F PET/CT test dataset. The results presented are average values.}
    \label{fig:extendeddata7}
\end{figure*}
\end{appendices}

\end{document}